%% file: ow2015.tex
\documentclass{article}
\usepackage{spconf,amsmath,graphicx}
\usepackage{amssymb}
\usepackage{mathtools}
\usepackage{graphicx}
\usepackage{epstopdf}
\usepackage{xargs}
\usepackage{ifthen}
\usepackage{bbm}
\usepackage{aliascnt}
\usepackage{stmaryrd}
\usepackage{todonotes}
\usepackage{enumerate}
\usepackage{ushort}
\usepackage{algpseudocode}
\usepackage{algorithm}
\usepackage{tikz}
\usetikzlibrary{positioning}

\numberwithin{equation}{section}
\input{defs}

\title{Efficient parameter inference in general hidden Markov models using the filter derivatives.}
\name{Jimmy Olsson\thanks{This work is supported by the Swedish Research Council, Grant 2011-5577.} and Johan Westerborn}
\address{KTH Royal Institute of Technology\\
Department of Mathematics\\
SE-100 44  Stockholm, Sweden}

\begin{document}
%
\maketitle
\begin{abstract}
Estimating online the parameters of general state-space hidden Markov models is a topic of importance in many scientific and engineering disciplines. In this paper we present an online parameter estimation algorithm obtained by casting our recently proposed particle-based, rapid incremental smoother (PaRIS) into the framework of recursive maximum likelihood estimation for general hidden Markov models. Previous such particle implementations suffer from either quadratic complexity in the number of particles or from the well-known degeneracy of the genealogical particle paths. By using the computational efficient and numerically stable PaRIS algorithm for estimating the needed prediction filter derivatives we obtain a fast algorithm with a computational complexity that grows only linearly with the number of particles. The efficiency and stability of the proposed algorithm are illustrated in a simulation study.
\end{abstract}
\section{Introduction}
\label{sec:intro}
This paper deals with the problem of \emph{online parameter estimation} in general state-space \emph{hidden Markov models} (HMMs) using \emph{sequential Monte Carlo} (SMC) methods and a \emph{recursive maximum likelihood} (RML) algorithm. HMMs with general state spaces, also referred to as \emph{state-space models}, are currently applied within a large variety of scientific and engineering disciplines, see, e.g.,~\cite[Chapter~1]{cappe:moulines:ryden:2005} and the references therein.

A \emph{hidden Markov model} (HMM) is a bivariate model consisting of an observable process $\{ Y_t \}_{t \in \nset}$, known as the \emph{observation process}, and an unobservable Markov chain $\{X_t\}_{t \in \nset}$, known as the \emph{state process}, taking values in some general state spaces $\set{Y}$ and $\set{X}$, respectively. We let $\xinit$ and $\hd{\parvec}$ denote the initial distribution and transition density (with respect to some reference measure denoted by $\rmd x$ for simplicity), respectively, of the hidden Markov chain, where $\parvec \in \Theta$ denotes a parameter vector that fully determines the dynamics of model and $\Theta$ is the parameter space. Conditioned on the state process, the observations are assumed to be independent with conditional distribution of $Y_t$ depending on $X_t$ only. We denote by $\md{\parvec}$ the density of the latter conditional distribution. Before using the HMM in practice, e.g. for prediction, the model needs to be calibrated through  estimation of the parameters $\theta$ given observations $y_{0:t} = (y_0, \ldots, y_t)$ (this will be our standard notation for vectors). There are several different approaches to this problem, based on either frequentist or Bayesian inference, and \cite{kantas:doucet:singh:maciejowski:chopin:2015} provides a good expos\'e of some current methods.

The \emph{maximum likelihood} approach aims at finding the parameter vector that maximizes the \emph{likelihood function} $\parvec \mapsto \llh{\parvec}(y_{0:t})$ or, equivalently, the \emph{log-likelihood function} $\parvec \mapsto \logllh{\parvec}(y_{0:t}) = \log \llh{\parvec}(y_{0:t})$. For the model specified above, the likelihood is given by 
\begin{multline*}
	\llh{\parvec}(y_{0:t}) \\
	= \int \md{\parvec}(x_0, y_0) \xinit(x_0) \prod_{s = 1}^t \md{\parvec}(x_s, y_s) \hd{\parvec}(x_{s - 1}, x_s) \, \rmd x_{0:t}.
\end{multline*}
The log-likelihood function and its gradient are generally intractable, as closed-form expressions are obtainable only in the linear Gaussian case or in the case where the state space $\set{X}$ is a finite set. Still, maximization of the log-likelihood function can be performed using the following \emph{Robbins-Monro scheme}: at iteration $n$, let $\parvec_n = \parvec_{n - 1} + \gamma_n Z_n$, where $Z_n$ is a noisy measurement of $\nabla \logllh{\parvec_{n - 1}}(y_{0:t})$, i.e. the gradient of the log-likelihood with respect to $\parvec$ evaluated at $\parvec = \parvec_{n - 1}$, and the sequence $\{ \gamma_n \}_{n \in \nset^*}$ of positive step sizes satisfies the regular stochastic approximation requirements $\sum_{n = 1}^\infty \gamma_n = \infty$ and $\sum_{n = 1}^\infty \gamma_n^2 < \infty$. 
Note that this approach requires the approximation $Z_n$ to be recomputed at every iteration of the algorithm. However, if the number of observations is very large, computing $Z_n$ is costly. Thus, since many iterations may generally be required for convergence, this yields an impractical algorithm. 

Hence, rather than incorporating the full observation record into each parameter update, it is preferable to update the parameters little by little as the data is processed. This is of course of particular importance in \emph{online} applications where the observations become available ``on-the-fly''. One such scheme is the following RML approach, where we instead update iteratively the parameters according to 
\begin{equation} \label{eq:rml:udate}
    \parvec_{t + 1} = \parvec_t + \gamma_{t + 1} \zeta_{t + 1} \eqsp, 
\end{equation}
where $\zeta_{t + 1}$ is a noisy observation of $\nabla \logllh{\parvec_t}(y_{t + 1} \mid y_{0:t})$, i.e. the gradient (with respect to $\theta$), evaluated at $\parvec = \parvec_t$, of the log-density of $Y_{t  + 1}$ given $Y_{0:t}$. This technique was, in the case of a finite state space $\set{X}$, studied in~\cite{legland:mevel:1997}. The same work also establishes, under certain conditions, the convergence of the output $\{ \parvec_t \}_{t \in \nset}$ towards the parameter $\parvec^*$ generating the data. 

In this note we present an efficient particle implementation of the previous RML scheme based on the \emph{particle-based, rapid incremental smoother} (PaRIS) introduced in~\cite{olsson:westerborn:2014} and analyzed further in~\cite{olsson:westerborn:2014b}. Compared with previous implementations, which suffer typically from quadratic complexity in the number of particles \cite{poyiadjis:doucet:singh:2005,delmoral:doucet:singh:2015}, our new algorithm is significantly faster as it has a computational complexity that grows only \emph{linearly} in the particle population size. The rest of the paper is organized as follows. After having introduced some basic notation in Section~\ref{sec:notation}, we delve deeper into the RML algorithm in Section~\ref{sec:recursive_maxmimum_likelihood}. In Section~\ref{sec:particle_methods} we present our new algorithm and finally we present, in Section~\ref{sec:simulations}, a simulation study, comparing our algorithm with an existing approach.

\section{Some notation}
\label{sec:notation}

In the following we let $\nset$ denote the set natural numbers and set $\nset^* = \nset \setminus \{0\}$. 

The conditional distribution of $X_{s:s'}$ given fixed observations $Y_{0:t} = y_{0:t}$ (with $(s, s') \in \nset^2$ and $s \leq s' \leq t$)  is given by 
\begin{multline*}
	\post{s:s' \mid t; \parvec}(x_{s:s'}) = \llh{\parvec}^{-1}(y_{0:t}) \iint \md{\parvec}(x_0, y_0) \xinit(x_0) \\ 
	\times \prod_{u = 1}^t \md{\parvec}(x_u, y_u) \hd{\parvec}(x_{u - 1}, x_u) \, \rmd x_{0:s - 1} \, \rmd x_{s' + 1:t} \eqsp.
\end{multline*}
We refer to $\post{t;\parvec} = \post{t \mid t; \parvec}$ and $\post{0:t \mid t; \parvec}$ as the \emph{filter} and \emph{joint smoothing distributions} at time $t$, respectively. The distribution
$$
\post{t + 1 \mid t; \parvec}(x_{t + 1}) = \int  \hd{\parvec}(x_t,  x_{t + 1}) \post{t + 1 \mid t; \parvec}(x_t) \, \rmd x_t,
$$
i.e., the conditional distribution of $X_{t + 1}$ given $Y_{0:t} = y_{0:t}$ is referred to as the \emph{prediction filter} at time $t$. In general, these distributions are intractable and need to be approximated. We will often consider expectations of some integrable function $\testf$ with respect to the posterior above and write $\post{s:s' \mid t ; \parvec}( \testf )= \int \testf(x_{s:s'}) \post{s:s' \mid t; \parvec}(x_{s:s'}) \, \rmd x_{s:s'}$.

For a vector of non-negative numbers $\{Êa_\ell \}_{i = 1}^N$ we let $\probdist(\{Êa_\ell \}_{i = 1}^N)$ denote the \emph{categorical distribution} induced by $\{Êa_\ell \}_{\ell = 1}^N$. In other words, $J \sim \probdist(\{Êa_\ell \}_{\ell = 1}^N)$ means that the random variable $J$ takes on the value $i$ with probability $a_i / \sum_{\ell = 1}^N a_\ell $.

\section{Recursive maximum likelihood} 
\label{sec:recursive_maxmimum_likelihood}

As mentioned in the introduction, the RML approach updates the parameter for each new observation $y_{t + 1}$ using \eqref{eq:rml:udate}. The algorithm that we propose is based on the fact that $\nabla \logllh{\parvec_t}(y_{t+1} \mid y_{0:t})$ can be decomposed as 
\begin{equation} 
    \begin{split}
    	\lefteqn{\nabla \logllh{\parvec_t}(y_{t + 1} \mid y_{0:t})} \\
    	&= \bigg( \int \post{t + 1 \mid t; \parvec_t}(x_{t + 1}) \nabla \md{\parvec_t}(x_{t + 1}, y_{t + 1}) \, \rmd x_{t + 1} \\
    	&+ \int \nabla \post{t + 1 \mid t; \parvec_t}(x_{t + 1}) \md{\parvec_t}(x_{t + 1}, y_{t + 1}) \, \rmd x_{t + 1} \bigg) \\
    	&\times \bigg( \int \post{t + 1 \mid t;\parvec_t}(x_{t + 1}) \md{\parvec_t}(x_{t + 1}, y_{t + 1}) \, \rmd x_{t + 1} \bigg)^{-1}, 
    \end{split} \label{eq:update}
\end{equation}
where $\nabla \post{t+1 \mid t; \parvec_t}(x_{t+1})$ is known as the \emph{prediction filter derivative} \cite{poyiadjis:doucet:singh:2005,delmoral:doucet:singh:2015} or \emph{tangent filter}. In order to compute the second integral in \eqref{eq:update} we follow the lines of~\cite[Section~2]{delmoral:doucet:singh:2015}; more specifically, write, for some $\post{t+1\mid t; \parvec}$-integrable function $\testf$,   
\begin{multline}
	\int \nabla \post{t+1 \mid t; \parvec}(x_{t+1}) \testf(x_{t+1}) \, \rmd x_{t+1} \\ = \E[\testf(X_{t + 1}) \Q{t + 1}(X_{0:t+1} ; \parvec) \mid y_{0:t}] \\
	- \E[\testf(X_{t + 1}) \mid y_{0:t}] \E[\Q{t+1}(X_{0:t+1} ; \parvec) \mid y_{0:t}], \label{eq:grad:joint}
\end{multline}
where $\Q{t}(x_{0:t} ; \parvec) = \sum_{s = 0}^{t - 1} \qsum{s}(x_{s:s + 1} ; \parvec)$, with
$$
    \qsum{s}(x_{s:s + 1} ; \parvec) = 
    \begin{cases}
        \nabla \log \left\{ \md{\parvec}(x_s, y_s) \hd{\parvec}(x_s, x_{s + 1}) \right\} & \mbox{for $s \in \nset^*$}, \\
        \nabla \log \md{\parvec}(x_0, y_0) & \mbox{for $s = 0$},
    \end{cases}
$$
is of additive form. In addition, using the tower property of conditional expectations, we may elaborate further \eqref{eq:grad:joint} according to
\begin{multline}
	\int \nabla \post{t+1 \mid t; \parvec}(x_{t+1}) \testf(x_{t+1}) \, \rmd x_{t+1} \\
	= \int \post{t+1 \mid t; \parvec}(x_{t+1}) \left\{ \tau_{t + 1}(x_{t + 1} ; \theta) \vphantom{\int} \right.  \\
	\left. - \int \tau_{t + 1}(x_{t + 1} ; \theta)  \post{t+1 \mid t; \parvec}(x_{t+1}) \, \rmd x_{t+1} \right\} f(x_{t+1}) \, \rmd x_{t+1}, 
        \label{eq:filt:deriv}
\end{multline}
with 
\begin{equation} \label{eq:tau:statistic}
    \tau_{t + 1}(x_{t + 1} ; \theta) = \E[\Q{t+1}(X_{0:t+1} ; \parvec) \mid X_{t + 1} = x_{t+1}, y_{0:t}]. 
\end{equation}
Combining the updating scheme \eqref{eq:rml:udate} with \eqref{eq:update} and \eqref{eq:filt:deriv}, with $\testf(x_{t+1})$ playing the role of $\md{\parvec_t}(x_{t+1},y_{t+1})$ in~\eqref{eq:filt:deriv}, gives us an RML algorithm, 
in which the noisy measurements $\{ \zeta_t \}_{t \in \nset^*}$  are produced via \eqref{eq:update} and \eqref{eq:filt:deriv} using a particle-based approach approximating online the prediction filter distributions as well as the statistics \eqref{eq:tau:statistic}. 

\section{PaRIS-based RML} 
\label{sec:particle_methods}

A \emph{particle filter} updates sequentially, using importance sampling and resampling techniques, a set $\{(\epart{t}{i}, \wgt{t}{i}) \}_{i = 1}^{\N}$ of particles and associated weights targeting a sequence of distributions. In our case we will use the particle filter to target the prediction filter flow $\{ \post{t + 1 \mid t; \parvec} \}_{t \in \nset}$ in the sense that for all $t \in \nset$ and $\post{t+1 \mid t; \parvec}$-integrable functions $f$, 
$$
	\post[part]{t+1 \mid t; \parvec}(\testf) = \frac{1}{\N} \sum_{i=1}^{\N} f(\epart{t+1}{i}) \backsimeq \post{t+1 \mid t; \parvec}(\testf) \quad \mbox{ as } \N \to \infty.
$$
While particle filters are generally well-suited for approximation of filter and prediction filter flows, estimation of the sequence $\{ \tau_t \}_{t \in \nset}$, where each $\tau_t$ is a conditional expectation over a \emph{path space}, is a considerably more challenging task. As established in several papers (see, e.g., \cite{olsson:westerborn:2014b,olsson:cappe:douc:moulines:2006,douc:garivier:moulines:olsson:2010}), standard sequential Monte Carlo methods fall generally short when used for path-space approximation due to particle ancestral lineage degeneracy. 
Since $\Q{t}(x_{0:t};\parvec)$ is additive, we express however the sequence $\{ \tau_t \}_{t \in \nset}$ \emph{recursively} as 
\begin{multline}
	\tau_{t + 1}(x_{t + 1}; \parvec) \\Ê
	= \E[ \tau_t(X_t; \parvec) + \qsum{t}(X_{t:t+1}; \parvec) \mid X_{t + 1} = x_{t+1}, y_{0:t}].
\label{eq:addf:update}
\end{multline}
Our aim is to approximate the recursion \eqref{eq:addf:update} on a grid of particles. Thus, letting, for $i \in \{1, \ldots, \N\}$, ${\tstattil{i}{t}}$ be approximations of $\tau_t(\epart{t}{i}; \parvec)$, we may approximate the recursion \eqref{eq:addf:update} by 
\begin{equation} \label{eq:slow:updates}
	\tstattil{i}{t+1} = \sum_{j = 1}^{\N} \frac{\wgt{t}{j} \hd{\parvec}(\epart{t}{j},\epart{t+1}{i})}{\sum_{\ell = 1}^{\N} \wgt{t}{\ell}\hd{\parvec}(\epart{t}{\ell},\epart{t+1}{i})} \left( \tstattil{j}{t} + \qsum{t}(\epart{t}{j}, \epart{t+1}{i}; \parvec) \right),
\end{equation}
where the ratio serves as a particle approximation of the \emph{backward kernel}, i.e., the conditional distribution of $X_t$ given $y_{0:t}$ and $X_{t + 1} = \epart{t + 1}{i}$; see~\cite{douc:garivier:moulines:olsson:2010,delmoral:doucet:singh:2009}. The recursion \eqref{eq:slow:updates} is the key component of the \emph{particle RML} proposed in \cite[Alg. 2]{delmoral:doucet:singh:2015}. However, since each update requires a sum of $\N$ terms to be computed, this approach has a computational complexity that grows quadratically with the the number of particles $\N$. 

As an alternative, the PaRIS algorithm \cite{olsson:westerborn:2014b} provides an efficient and numerically stable estimator of smoothed expectations of additive form and hence a more convenient way of approximating the updating formula~\eqref{eq:addf:update}. More specifically, we replace each update \eqref{eq:slow:updates} by the PaRIS-type update 
\begin{equation*}
	\tstat[i]{t+1} = \K^{-1} \sum_{\k = 1}^{\K} \left( \tstat[\bi{t+1}{i}{\k}]{t} + \qsum{t}(\epart{t}{\bi{t+1}{i}{\k}}, \epart{t+1}{i} ; \parvec) \right),
\end{equation*}
where $\K \in \nset^*$ is typically small and $\{ \bi{t+1}{i}{\k} \}_{\k = 1}^{\K}$ are i.i.d. indices drawn from $\probdist( \{ \wgt{t}{\ell} q(\epart{t}{\ell}, \epart{t}{i} ) \}_{\ell = 1}^\N )$. Using an accept-reject sampling approach proposed in \cite{douc:garivier:moulines:olsson:2010}, each index can be drawn efficiently with a cost that can be proven to be uniformly bounded in $\N$ and $t$. This yields an algorithm with complexity $\mathcal{O}(\N \K)$ instead of $\mathcal{O}(\N^2)$. A key discovery made in \cite{olsson:westerborn:2014b} is that the PaRIS algorithm converges, as $\N$ tends to infinity, for all \emph{fixed} $\K \in \nset^*$ and stays numerically stable, as $t$ tends to infinity, when $\K \geq 2$; in particular, it is, for large $\N$ and when $\K \geq 2$, possible to derive a time-uniform $\mathcal{O}(\{1 + 1/(\K - 1)\}/\N)$ bound on the variance of the PaRIS estimator, suggesting that $\K$ should be kept at a \emph{moderate value} (as increasing $\K$ will not lead to significant reduction of variance). In fact, using only $\K = 2$ draws has turned out to work well in simulations (see \cite[Section~4]{olsson:westerborn:2014b}). We refer to  \cite{olsson:westerborn:2014b} for a detailed discussion. Compared to the $\mathcal{O}(\N^2)$ algorithm based on the updates \eqref{eq:slow:updates}, the extra randomness introduced by the PaRIS-based algorithm adds some variance to the estimates; still, it will be clear in the simulations that follow that the  linear computational complexity of our approach allows considerably more particles to be used, resulting in a significantly smaller variance for a given computational budget. 

\subsection{PaRIS-based RML algorithm} 
\label{ssub:rml_algorithm}

In the algorithm below, all operations involving the index $i$ should be performed for all $i \in \{1, \ldots, \N\}$. The precision parameter $\K$ is to be set by the user. In addition $\term{t}{1}$, $\term{t}{2}$ and $\term{t}{3}$ correspond to the three terms of \eqref{eq:update}.
\begin{algorithmic}[1]
\State Initialize the algorithm by setting arbitrarily $\parvec_0$
\State Draw $\epart{0}{i} \sim \xinit$
\State Set $\tstat[i]{0} \gets 0$
\For{$t \gets 0, 1, 2, \ldots$}
\State Set $\wgt{t}{i} \gets \md{\parvec_t}(\epart{t}{i}, y_t)$
\State Draw $I_i \sim \probdist(\{\wgt{t}{i}\}_{i=1}^{\N})$
\State Draw $\epart{t + 1}{i} \sim \hd{\parvec_t}(\epart{t}{I_i}, \cdot)$
\For{$\k \gets 1, \ldots, \K$}
\State Draw $\bi{t + 1}{i}{\k} \sim \probdist( \{ \wgt{t}{\ell} \hd{\parvec_t}(\epart{t}{\ell}, \epart{t + 1}{i} ) \}_{\ell = 1}^\N)$
\EndFor
\State Set 
$$
\tstat[i]{t + 1} \gets \K^{-1}\sum_{\k = 1}^{\K}\left( \tstat[\bi{t + 1}{i}{\k}]{t} + \qsum{t}(\epart{t}{\bi{t + 1}{i}{\k}}, \epart{t + 1}{i} ; \parvec_t) \right)
$$
\State Set $\bar{\tau}_{t + 1} \gets \N^{-1} \sum_{\ell = 1}^{\N} \tstat[\ell]{t + 1}$
\State Set	$\term{t + 1}{1} \gets \N^{-1} \sum_{\ell = 1}^{\N} \nabla \md{\parvec_t}(\epart{t + 1}{\ell}, y_{t + 1})$
\State Set $\term{t + 1}{2} \gets \N^{-1} \sum_{\ell = 1}^{\N} \left( \tstat[\ell]{t + 1} - \bar{\tau}_{t + 1} \right) \md{\parvec_t}(\epart{t + 1}{\ell}, y_{t + 1})$
\State Set $\term{t + 1}{3} \gets \N^{-1} \sum_{\ell = 1}^{\N} \md{\parvec_t}(\epart{t + 1}{\ell}, y_{t + 1})$
\State Set $\parvec_{t + 1} \gets \parvec_t + \gamma_{t + 1} \dfrac{\term{t + 1}{1} + \term{t + 1}{2}}{\term{t + 1}{3}}$
\EndFor
\end{algorithmic}

\begin{figure*}
	\begin{minipage}[b]{0.48\linewidth}
		\centering
		\centerline{\includegraphics[width=8.5cm]{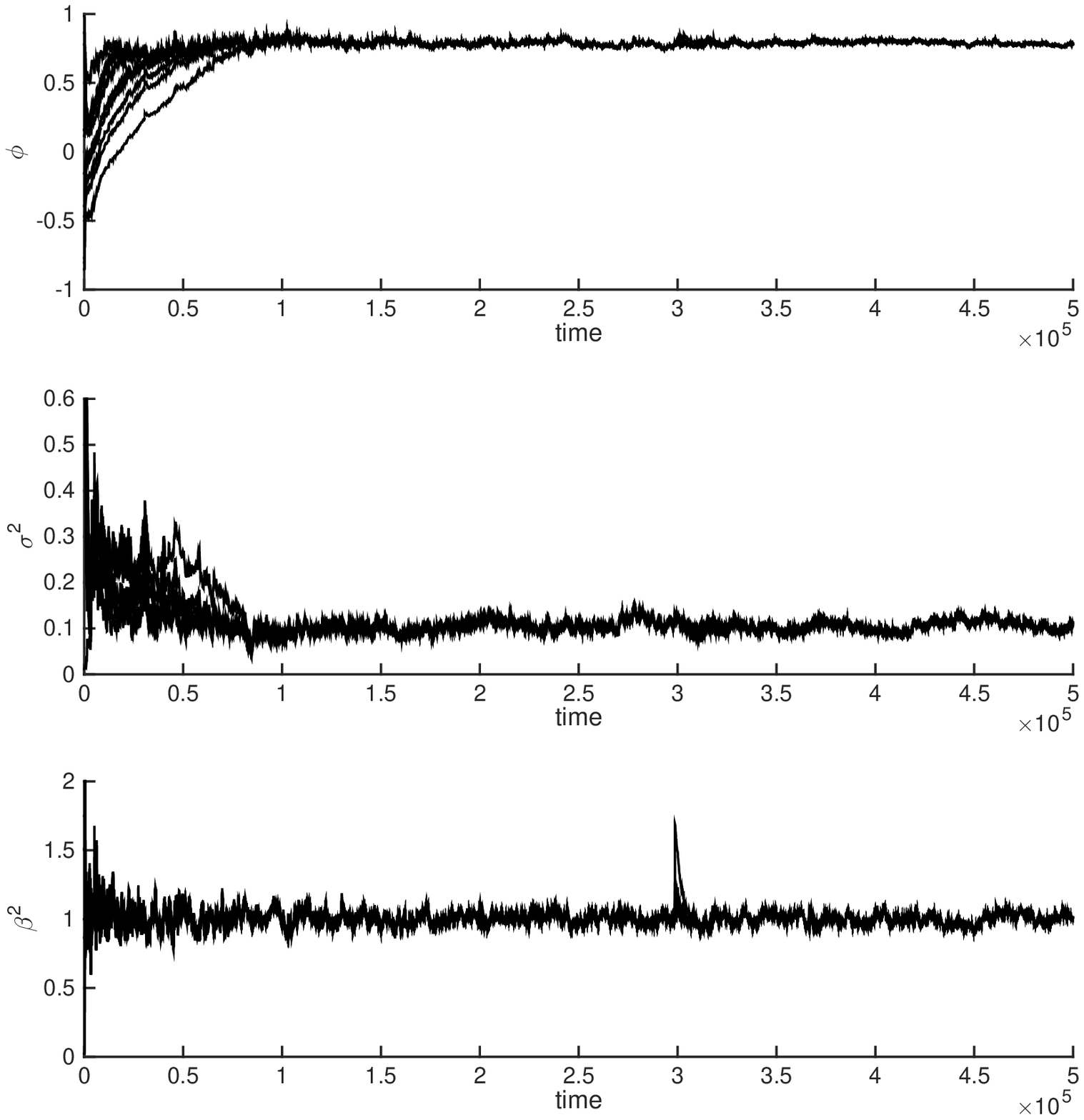}}
		\centerline{(a) PaRIS-based RML}\medskip
	\end{minipage}
	\begin{minipage}[b]{0.48\linewidth}
		\centering
		\centerline{\includegraphics[width=8.5cm]{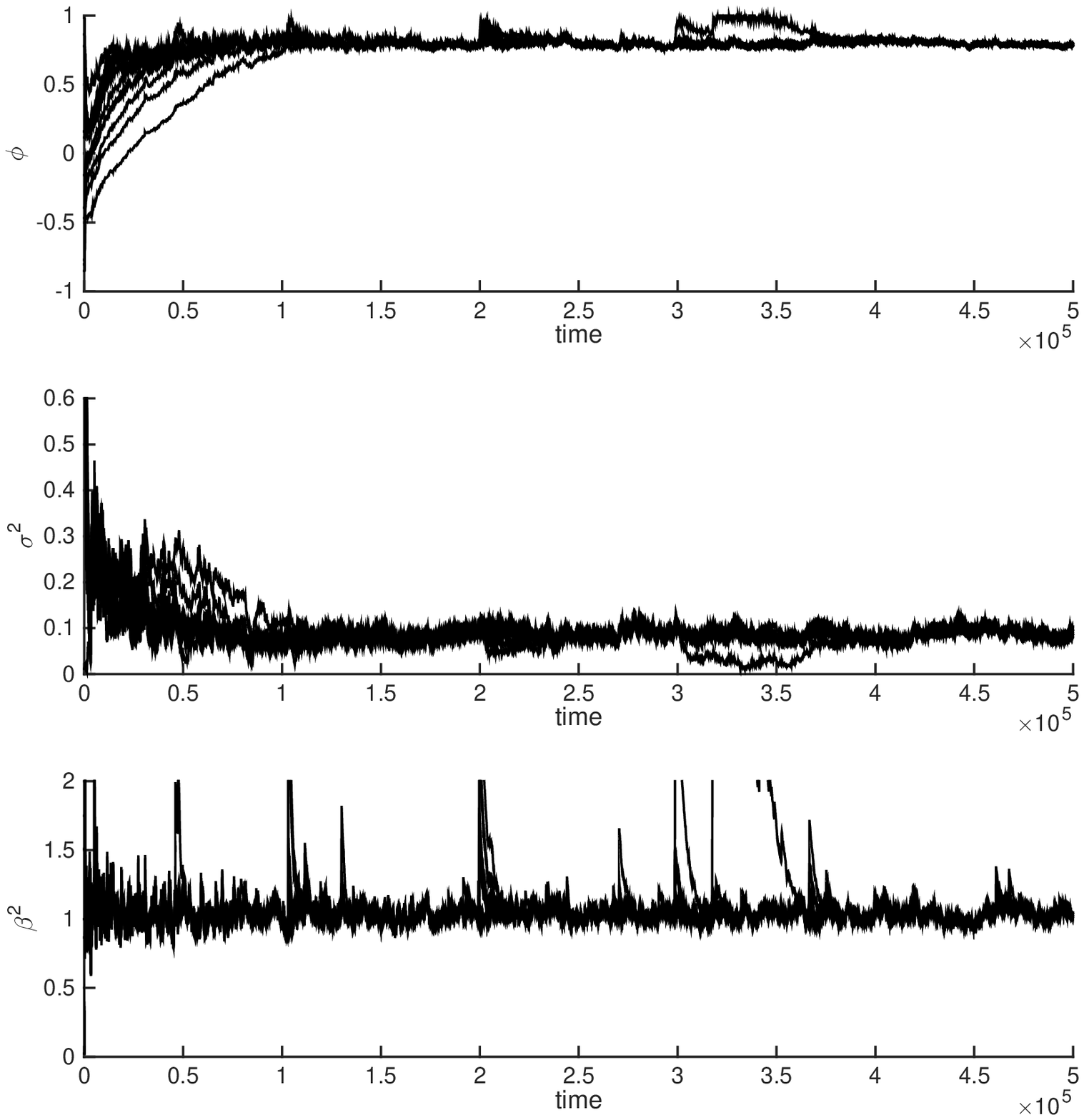}}
		\centerline{(b) Particle RML}\medskip
	\end{minipage}
	\caption{Particle learning trajectories produced by the PaRIS-based (left panel) and standard particle (right-panel) RML for, from top to bottom, $\phi$, $\sigma^2$ and $\beta^2$. For each algorithm, 12 replicates were generated on the same data set with different, randomized initial parameters (being the same for both algorithms). For the particle RML, the plot of $\beta^2$ does not contain the full trajectories due to very high peaks.}
	\label{fig:trajectories}
\end{figure*}

\section{Simulations} 
\label{sec:simulations}
We tested our method on the stochastic volatility model 
\[
    \begin{split}
        X_{t + 1} &= \phi X_t + \sigma V_{t + 1} \eqsp, \\
        Y_t &= \beta \exp(X_t / 2) U_t \eqsp,
    \end{split}
    \quad t \in \nset \eqsp,
\]
where $\{V_t \}_{t \in \nset^*}$ and $\{U_t\}_{t \in \nset}$ are independent sequences of mutually independent standard Gaussian noise variables. Parameters to be estimated were $\parvec = (\phi, \sigma^2, \beta^2)$, and we compared the performance of our PaRIS-based RML to that of the particle RML proposed in \cite{delmoral:doucet:singh:2015}. To get a fair comparison of the algorithms we set the number of particles used in each algorithm such that both algorithms ran in the same computational time. With our implementation, $\N = 100$ for the particle RML corresponded to $\N = 1400$ and $\K = 2$ for the PaRIS-based RML. For both algorithms we set $\gamma_t = t^{-0.6}$. The algorithms were executed on data comprising $500\,000$ observations generated under the parameter $\parvec^* = (0.8, 0.1,1)$. Each algorithm ran $12$ times with the same observation input but with randomized starting parameters (still, the same starting parameters were used for both algorithms). In Fig.~\ref{fig:trajectories} we present the resulting learning trajectories, and it can clearly be seen that the PaRIS-based RML exhibits significantly less variance in its estimates, especially for the $\beta^2$ variable. In the particle RML we notice some large jumps in the $\beta^2$ variable, originating from the fact that the corresponding estimate of $\term{t}{3}$ gets very small. This is due to the low number of particles failing to cover the support of the emission density. In contrast, since we can utilize considerably more particles in the PaRIS-based RML, we see only a single, comparably small, jump in $\beta^2$. Judging by the estimated (on the basis of the $12$ trajectories) variances $(.054, .164, .063 ) \times 10^{-3}$ and $(.069, .181, .095) \times 10^{-4}$  of the final parameter estimates for the particle RML and the PaRIS-based RML, respectively, the PaRIS-based RML is roughly ten times more precise than the particle RML.

\section{Discussion} 
\label{sec:summary}
We have proposed a novel algorithm for online parameter learning in general HMMs using an RML method based on the PaRIS algorithm \cite{olsson:westerborn:2014b}. The new method has a linear computational complexity in the number of particles, which allows considerably more particles to be used for a given computational budget compared to previous methods. The performance of the algorithm is illustrated by simulations indicating clearly improved convergence properties of the parameter estimates. 



\bibliographystyle{IEEEbib}
\bibliography{biblio}

\end{document}

%% file: defs.tex
\newcommand{\bi}[3]{J_{#1}^{(#2, #3)}}
\newcommandx{\bk}[2][1=]{ 
\ifthenelse{\equal{#1}{}}
{\overleftarrow{q}_{#2}}
{\overleftarrow{\kernel{Q}}_{#2}^{#1}}
}

\newcommand{\E}{\mathbb{E}}
\newcommand{\epart}[2]{\xi_{#1}^{#2}}

\newcommand{\eqsp}{}

\newcommandx\filtd[2][1=]{
\ifthenelse{\equal{#1}{}}
	{\varphi_{#2}}
	{\varphi_{#2}^\N}
}

\newcommand{\hd}[1]{q_{#1}}

\renewcommand{\k}{j}
\newcommand{\K}{\tilde \N}
\newcommand{\kernel}[1]{\mathbf{#1}}

\newcommand{\llh}[1]{L_{#1}}
\newcommand{\logllh}[1]{\ell_{#1}}

\newcommand{\N}{N}
\newcommand{\nset}{\mathbb{N}}

\newcommand{\md}[1]{g_{#1}}

\newcommand{\parvec}{\theta}
\newcommandx\post[2][1=]{
\ifthenelse{\equal{#1}{}}
	{\phi_{#2}}
	{\phi_{#2}^\N}
}
\newcommand{\probdist}{\mathsf{Pr}}

\newcommand{\Q}[1]{h_{#1}}
\newcommand{\qsum}[1]{\tilde{h}_{#1}}

\newcommand{\rmd}{\mathrm{d}}

\newcommand{\set}[1]{\mathsf{#1}} 

\newcommand{\term}[2]{\zeta_{#1}^{#2}}
\newcommand{\testf}{f}
\newcommand{\tstatletter}{\kernel{T}}
\newcommandx\tstat[2][1=]{
\ifthenelse{\equal{#1}{}}
	{\tstatletter_{#2}}
	{\tau_{#2}^{#1}}
}
\newcommand{\tstattil}[2]{\tilde{\tau}_{#2}^{#1}}

\newcommand{\wgt}[2]{\omega_{#1}^{#2}}

\newcommand{\xinit}{\chi}

%% file: ow2015.bbl
\begin{thebibliography}{10}

\bibitem{cappe:moulines:ryden:2005}
O.~Capp\'{e}, E.~Moulines, and T.~Ryd\'{e}n,
\newblock {\em Inference in Hidden {M}arkov Models},
\newblock Springer, 2005.

\bibitem{kantas:doucet:singh:maciejowski:chopin:2015}
N.~Kantas, A.~Doucet, S.~S. Singh, J.~Maciejowski, and N.~Chopin,
\newblock ``On particle methods for parameter estimation in state-space
  models,''
\newblock {\em Statist. Sci.}, vol. 30, no. 3, pp. 328--351, 08 2015.

\bibitem{legland:mevel:1997}
F.~{Le~Gland} and L.~Mevel,
\newblock ``Recursive estimation in {HMM}s,''
\newblock in {\em Proc. IEEE Conf. Decis. Control}, 1997, pp. 3468--3473.

\bibitem{olsson:westerborn:2014}
J.~Olsson and J.~Westerborn,
\newblock ``Efficient particle-based online smoothing in general hidden
  {M}arkov models,''
\newblock in {\em IEEE 2014 International Conference on Acoustics, Speech, and
  Signal Processing (ICASSP 2014)}, 2014.

\bibitem{olsson:westerborn:2014b}
J.~{Olsson} and J.~{Westerborn},
\newblock ``{Efficient particle-based online smoothing in general hidden Markov
  models: the PaRIS algorithm},''
\newblock {\em Bernoulli}, 2016,
\newblock to appear.

\bibitem{poyiadjis:doucet:singh:2005}
G.~Poyiadjis, A.~Doucet, and S.~S. Singh,
\newblock ``Particle methods for optimal filter derivative: application to
  parameter estimation,''
\newblock in {\em Proc. IEEE Int. Conf. Acoust., Speech, Signal Process.},
  18-23 March 2005, pp. v/925--v/928.

\bibitem{delmoral:doucet:singh:2015}
P.~Del~Moral, A.~Doucet, and S.~S. Singh,
\newblock ``Uniform stability of a particle approximation of the optimal filter
  derivative,''
\newblock {\em SIAM Journal on Control and Optimization}, vol. 53, no. 3, pp.
  1278--1304, 2015.

\bibitem{olsson:cappe:douc:moulines:2006}
J.~Olsson, O.~Capp\'e, R.~Douc, and E.~Moulines,
\newblock ``Sequential {M}onte {C}arlo smoothing with application to parameter
  estimation in non-linear state space models,''
\newblock {\em Bernoulli}, vol. 14, no. 1, pp. 155--179, 2008.

\bibitem{douc:garivier:moulines:olsson:2010}
R.~Douc, A.~Garivier, E.~Moulines, and J.~Olsson,
\newblock ``Sequential {M}onte {C}arlo smoothing for general state space hidden
  {M}arkov models,''
\newblock {\em Ann. Appl. Probab.}, vol. 21, no. 6, pp. 2109--2145, 2011.

\bibitem{delmoral:doucet:singh:2009}
P.~Del~Moral, A.~Doucet, and S.~Singh,
\newblock ``Forward smoothing using sequential {M}onte {C}arlo,''
\newblock Tech. {R}ep., Cambridge University, 2010.

\end{thebibliography}
